# Chiral Properties of Bismuth Ferrite (BiFeO$_3$) Inferred from Resonant x-ray Bragg Diffraction

Angel Rodriguez-Fernandez [1]*, Stephen William Lovesey [2,3], Steve Patrick Collins [3], Gareth Nisbet [3] and Jesus Angel Blanco [1]

[1] *Physics Department, University of Oviedo, C/ Calvo Sotelo s/n Oviedo, Spain*
[2] *ISIS Facility, STFC Oxfordshire OX11 0QX, UK*
[3] *Diamond Light Source Ltd, Oxfordshire OX11 0DE, UK*

A new chiral phase of ferric ions in bismuth ferrite, the only material known to support multiferroic behaviour at room temperature, is inferred from extensive sets of data gathered by resonant x-ray Bragg diffraction. Values of all ferric multipoles participating in a minimal model of Fe electronic structure are deduced from azimuthal-angle scans. Extensive sets of azimuthal-angle data, gathered by resonant x-ray Bragg diffraction, yield values of all ferric multipoles participating in a minimal model of Fe electronic structure. Paramagnetic (700 K) and magnetically ordered (300 K) phases of a single crystal of BiFeO$_3$ have been studied with x-rays tuned near to the iron K-edge (7.1135 keV). At both temperatures, intensities at a Bragg spot forbidden in the nominal space-group, R3c, are consistent with a chiral motif of ferric ions in a circular cycloid propagating along $(1, 1, 0)_H$. Templeton and Templeton scattering at 700 K is attributed in part to charge-like quadrupoles in a cycloid. The contribution is not present in a standard, simplified model of electronic states of the resonant ion with trivial cylindrical symmetry.

KEYWORDS: multiferroic state, chirality, resonant Bragg diffraction

An electronic state in which charge and magnetic polarizations coexist has been at the centre of materials science in the past decade. Such a state exists in bismuth ferrite (BiFeO$_3$) at room temperature, while in all other known cases a multiferroic state emerges upon cooling. This makes bismuth ferrite a unique candidate for potential application in electronic devices, such as sensors or multi-state memory storage-units. [1-4] We have gathered evidence for a new chiral phase of BiFeO3 in the paramagnetic phase using resonant x-ray Bragg diffraction and, by way of a test for the chirality property, we demonstrate that our proposed electronic structure allows coupling to radiation with a like property, namely, x-rays with circular polarization (helicity).

Moreover, our extensive sets of diffraction data enable us to infer values of ferric multipoles

---

*E-mail: angelrf86@gmail.com

in both the paramagnetic and magnetically ordered phases.

Bismuth ferrite forms in a rhombohedrally distorted perovskite crystal structure of R3c-type (#161). Weak ferroelectricity develops below a Curie temperature $T_c \approx$ 1100 K, and G-type antiferromagnetic order of ferric ($Fe^{3+}$) dipole moments is observed below a Néel temperature $T_N \approx$ 640 K. The antiferromagnetism coexists with a long-period modulation ($\approx$ 620 Å) in the hexagonal plane, as shown in figure 1.[5,6] This coexistence is a curious property of a quite simple compound, created by the tension between two interactions that favor parallel and orthogonal spin arrangement, respectively.

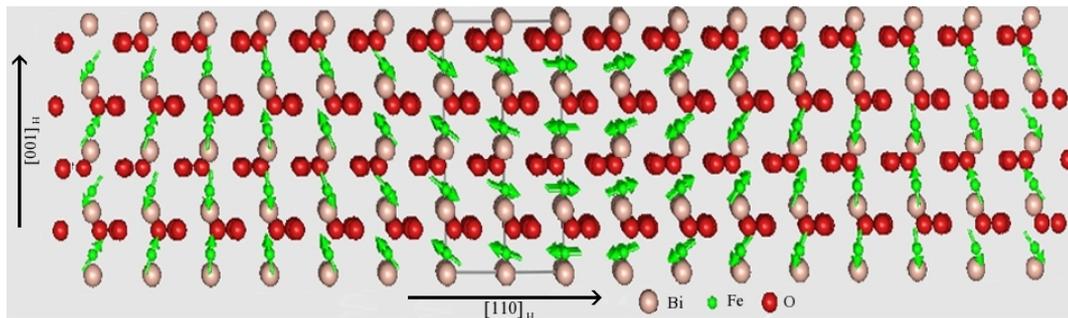

Fig. 1. Scheme of the crystal and magnetic structures of bismuth ferrite ($BiFeO_3$), with hexagonal setting. The $[0,0,1]_H$ axis is vertical. Directions of magnetic dipoles of the Fe ions at room temperature are indicated by arrows.

For studies of electronic magnetism, the experimental technique of resonant x-ray Bragg diffraction we use has advantages over non-resonant diffraction that is difficult to exploit quantitatively, because uncertainty surrounds the deployment of a scattering length asymptotically valid in the Compton region of scattering.[7-11] Resonant x-ray diffraction has proved its worth in many studies, particularly those that focus on one or more of the raft of electronic properties driven by angular anisotropy in valence states. [12-13] Intensities collected at weak, space-group forbidden reflections access directly information about complex electronic structure manifest in atomic multipoles, including, magnetic charge (or magnetic monopole),[14] electric dipole,[15] anapole,[16,17] quadrupole,[18] octupoles,[19,20] and hexadecapoles.[21,22] In consequence, weak reflections are extremely sensitive to charge, orbital and spin electron degrees of freedom and $BiFeO_3$, with a chemical structure similar to haematite, is no exception.[23]

Crystal growth was performed in platinum crucibles with content of about 90 g, using the accelerated rotation technique, and a platinum cover welded tightly to the crucible, leaving only a central hole of 0.1 mm diameter, as explained in reference.[24] The size of the sample was about 5 x 5mm$^2$ and a thickness of 0.5 mm, showing a polished surface in the [0, 0, l]$_H$ direction.

*E-mail: angelrf86@gmail.com

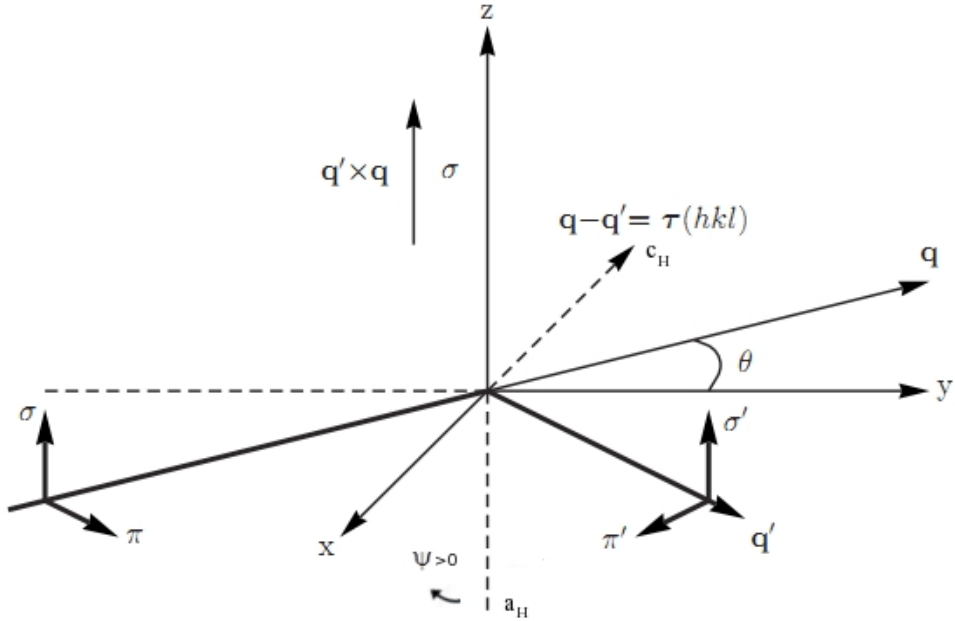

Fig. 2. Cartesian coordinates (x, y, z) and x-ray polarization and wavevectors. The plane of scattering spanned by primary (**q**) and secondary (**q'**) wavevectors coincides with the x-y plane. Polarization labelled σ and σ' is normal to the plane and parallel to the z-axis, and polarization labelled π and π' lies in the plane of scattering. The beam is deflected through an angle 2θ. Nominal setting of the crystal is indicated with $a_h$
 antiparallel to σ-polarization, together with sense of rotation in an azimuthal-angle

Our hexagonal crystal coordinates are $a_h$ = a(1, 0, 0), $b_h$ = a(–1/2, √3/2, 0) and $c_h$ = c(0, 0, 1), with a = 5.58Å and c = 13.88Å.[25] Basis vectors, or principal crystal axes, are ξ = (1, 0, 0), η = (0, 1, 0) and ζ = (0, 0, 1), and they coincide with (x, y, z) in figure 2 at the nominal setting of the crystal. The Bragg wavevector $(0, 0, l)_H$ is aligned with −x, as shown in figure 2. Intensities are measured as a function of rotation of the crystal about the Bragg wavevector through an angle ψ.

The $(0, 0, 9)_H$ reflection is forbidden in the nominal space group R3c. Bragg diffraction due to angular anisotropy in available valence states is weak but, none the less, visible in diffraction enhanced by an atomic resonance, as evident in data displayed in figure 3. Resonant x-ray diffraction experiments were performed at the Diamond Light Source (UK), on beamline I16. The horizontally polarized beam, σ, was tuned near the iron K-edge (7.1135 keV). We observed intensity at the $(0, 0, 9)_H$ reflection in two studies with the sample held at a temperature below (300 K) and above (700 K) the Néel temperature. The change in intensity that we observed with cooling, between the two temperatures, confirmed the magnetic origin of the difference signal; relevant data are displayed in Figure 5. All data were collected in the rotated channel of polarization π'σ, where states of polarization labelled π' and σ are defined in figure 2. During the experiment we scan the surface of the sample to try to determine the size of the domains, and select the appropriate region of the sample where a likely single domain could be involved in the scattering process. While it is possible that the sample supports different domains, the present results were consistent with a single domain illuminated by the very small size of the primary beam (180 x 40 μm$^2$).

*E-mail: angelrf86@gmail.com

The azimuthal scans presented in Figure 4 were obtained performing "θ scans" with the detector around the Bragg condition for different azimuthal angles. This method was used previously by Finkelstein et al.[19] and by Kokubun et al.[44], among others. The experimental values displayed in Figures 4 and 5 are the integrated intensity of each of the curves normalized to intensity in the primary x-ray beam. Due to the small penetration depth of an allowed Bragg spot (0, 0, 6)$_H$, where the diffraction is mostly following a dynamical process, we have not considered to use this kind of reflection to normalize a forbidden Bragg spot (0, 0, 9)$_H$ that, due to its weakness has a kinematical behaviour a larger penetration depth and less affected by possible defects from the surface.

All reasonable steps have been taken to arrive at sound data. Subtraction of the background intensity due to Renninger reflections (multi-beam peaks), observed in an azimuthal scan, was done using a Matlab program available at the instrument. For the case of room temperature, an azimuthal scan was done to select optimum, flat positions between peaks and avoid the Renninger effect (therefore measured points in the azimuth dependence are not equidistant). Due to the fact that we have collected resonant x-ray data for a certain selected reflection at different parts of the single crystal, we consider that the experimental data shown in figures 4 & 5 are related to the resonant event rather than to the tail of the Renninger effect. The high-quality crystal used for the experiment has a face perpendicular to the (0, 0, 1)$_H$ direction, so the experiment was performed with a specular geometry.

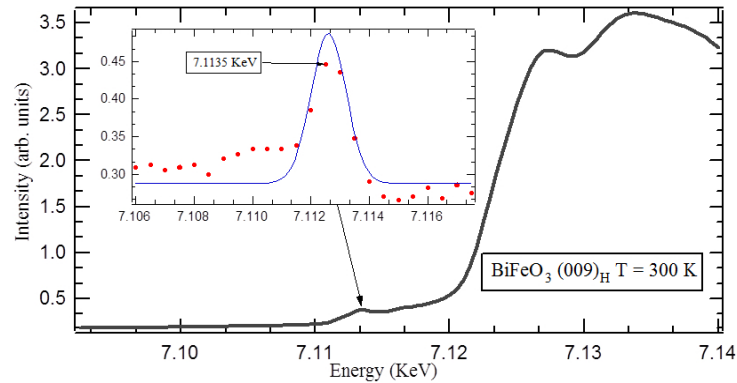

Fig. 3. (Colour Online) X-ray spectrum in the vicinity of the Fe K-edge for the (009)$_H$ reflection. Diffraction data reported in figures 4 and 5 were collected tuning the primary energy to E = 7.1135 keV. Inset: (Red dots) Energy scan data and (blue line) approximation to a harmonic oscillator.

*E-mail: angelrf86@gmail.com

We address, first, Templeton and Templeton (T & T) scattering reported in figure 4 (filled dots) measured with the sample at 700 K ($T_N \approx$ 640 K).[26] Resonant x-ray diffraction enhanced by an electric dipole - electric dipole (E1-E1) event is forbidden at the (0, 0, 9)$_H$ Bragg spot of a R3c-type chemical structure. Diffraction enhanced by an electric quadrupole-electric quadrupole (E2-E2) event is allowed, however, and it is produced by an electric, time-even hexadecapole. This diffraction is part of what we have observed, as we now explain.

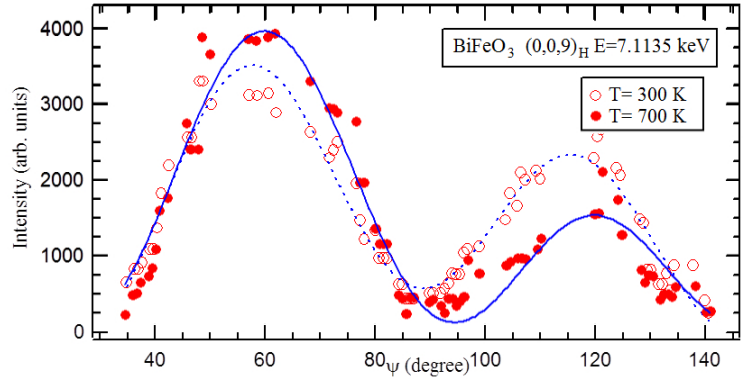

Fig. 4. (full dots) Intensity of the Bragg spot (0, 0, 9)$_H$ as a function of azimuthal angle, $\psi$, with a sample temperature of 700 K, forbidden in the R3c-type structure and called Templeton and Templeton (T & T) scattering. Rotation of the crystal is counter clockwise about the Bragg wavevector, and the origin $\psi = 0$ is $\mathbf{a}_h$ antiparallel to σ-polarization, figure 2. (empty dots) Intensity as a function of azimuthal angle obtained at room temperature, 300K. Corrections to raw data are described in the text. Solid (dashed) line is a fit to our model of diffraction by a motif of charge-like quadrupoles and hexadecapoles, namely, $|F_{\pi'\sigma}|^2$ and expression (6) with v = 0 (v ≠ 0).

Our notation for parity-even atomic multipoles is $\langle T^K_Q \rangle$, with a complex conjugate $\langle T^K_Q \rangle^* = (-1)^Q \langle T^K_{-Q} \rangle$, where the positive integer K is the rank and Q the projection, with $-K \leq Q \leq K$. Angular brackets $\langle ... \rangle$ denote the time average of the enclosed spherical tensor operator, i.e., multipoles are properties of the electronic ground-state, and the time-signature of $\langle T^K_Q \rangle$ is $(-1)^K$.[12,13] The hexadecapole (K = 4) in question is the real part of $\langle T^4_{+3} \rangle$, denoted by $\langle T^4_{+3} \rangle'$. A triad axis of symmetry, $C_{3z}$, passes through the iron sites, 6a in R3c. Diffraction by these ions using a Bragg wavevector (0, 0, l)$_H$ is described by an electronic structure factor,

$$\Psi^K_Q = \{1 + 2\cos(2\pi l/3)\} [\langle T^K_Q \rangle + (-1)^l (-1)^K \langle T^K_{-Q} \rangle]. \tag{1}$$

Space-group allowed reflection are defined by diagonal elements $\Psi^K_0$ with K even, and $\Psi^K_0 \neq 0$ is allowed for $l = 6n$. The identity $C_{3z} \langle T^K_Q \rangle = \langle T^K_Q \rangle$ requires Q = ± 3m. As anticipated, E1-E1 is forbidden for $l$ odd, because, of course, $\Psi^K_0 = 0$ for a space-group forbidden reflection, while Q = ± 3 does not contribute to a dipole-dipole event where K does not exceed 2. Using (1) for an E2-E2 event, we find the corresponding unit-cell structure

---

*E-mail: angelrf86@gmail.com

factor is a three-fold periodic function of the azimuthal angle, $\psi$,[27]

$$F_{\pi'\sigma} = (3/\sqrt{2})\cos^3\theta \cos(3\psi) \langle T^4_{+3}\rangle'. \qquad (2)$$

In this expression, $\theta$ is the Bragg angle, and $\mathbf{a}_h$ is antiparallel to $\sigma$-polarization, normal to the plane of scattering in figure 2, at the origin of an azimuthal-angle scan, $\psi = 0$. Intensity corresponding to (2), $|F_{\pi'\sigma}|^2 \propto \cos^2(3\psi)$, is symmetric about $\psi = 90°$, which does not agree with our data for T & T scattering displayed in figure 4 (filled dots).

Missing in what has been described thus far, we propose, is T & T scattering caused by charge-like quadrupoles (K = 2) in a circular cycloid, using an E1-E1 event. An electric dipole (E1) is expected to be appreciably stronger than an electric quadrupole (E2) event. But diffraction from the quadrupoles is weak, being the responsibility of components absent in a standard stick-model, in which the electronic state of the resonant ion is restricted to cylindrical symmetry.[28] Whence, the minimal model that explains measurements in figure 4 (700 K, filled dots) is a sum of two forms of weak T & T scattering. Adding the corresponding magnetic scattering, we achieve a model that explains data displayed in figure 4 (300 K, empty dots).

We invoke a (circular) cycloid with the plane of the cycloid parallel to the plane spanned by $\mathbf{c}_h$ and $\mathbf{a}_h + \mathbf{b}_h$. This motif is one candidate considered by Przeniosło et al; see Model 1 in figure 1.[29] We will assume that the cycloid, composed of charge-like multipoles, is constant, independent of temperature, to a good approximation. This is a sound assumption for the paramagnetic phase, and not unreasonable at lower temperatures for multipoles not induced by magnetic order.

Starting with an explanation of T & T scattering in figure 4, we utilize quadrupoles for a circular cycloid, $\langle C^2_Q\rangle$, introduced by Scagnoli and Lovesey[27] and recently reviewed by Lovesey et al. [45] These quadrupoles, in common with all cycloid multipoles, are not subject to the symmetry operations in the point group for sites 6a in the R3c group. In the general case, one finds $\langle C^2_0\rangle = 0$ for the first harmonic of the cycloid, which is the one of interest. For a circular cycloid rotating in the x-z plane,

$$\langle C^2_{+1}\rangle = (1/4)[\langle T^2_{+1} + T^2_{-1}\rangle + i\langle T^2_{+2} - T^2_{-2}\rangle] \equiv -(1/\sqrt{6})[i(yz) + (xz)], \qquad (3)$$

where $(\alpha\beta)$ is a standard, traceless second-rank Cartesian tensor. A representation of the quadrupole, $\langle \mathbf{T}^2\rangle$, in terms of standard operators is available.[30]

By way of an orientation to the result (3) we consider its value for a standard stick-model.[28] In this case, all electronic properties of the resonant ion are manufactured from one material vector. Using $\alpha$ and $\beta$ to represent Cartesian coordinates, a general second-rank tensor $(\alpha\beta)$ is to be replaced by a simple product $\langle\alpha\rangle\langle\beta\rangle$, leading to $(yz) = (xy) = 0$ for a material vector confined to the x-z plane with ⟨y⟩ = 0.

*E-mail: angelrf86@gmail.com

Guided by R3c, we need the quadrupole (3) and the quadrupole derived from it by rotation about ($a_h$ + $b_h$) by 180°. The sum of the two correctly related quadrupoles is transformed to principal crystal-axes with the result,

$$\Psi^2_{+1} = \sqrt{(3/2)} \,[(\sqrt{3}/4) \{(\xi^2 - \eta^2) + 3(\zeta\zeta)\} + (\eta\zeta)], \qquad (4)$$

where Cartesian quadrupoles are referred to previously defined principal crystal-axes. Note that $\Psi^2_{+1}$ is purely real. Turning to data obtained with a sample temperature of 300 K and displayed in figure 4 (empty dots), magnetic diffraction by the cycloid is created by a time-odd dipole,

$$\langle C^1_0 \rangle = (1/2)\,[\langle T^1_0 \rangle + i\,\langle T^1_{+1} - T^1_{-1}\rangle/\sqrt{2}]. \qquad (5)$$

For a reflection (0, 0, l)$_H$ with l odd, it actually contributes a magnetic dipole parallel to ($a_h$ + $b_h$), namely, ⟨$T^1_{-1}$ - $T^1_{+1}$⟩/$\sqrt{2}$ calculated with principal crystal axes. At the K-edge, a dipole ⟨$T^1$⟩ is simply orbital angular momentum.[31]

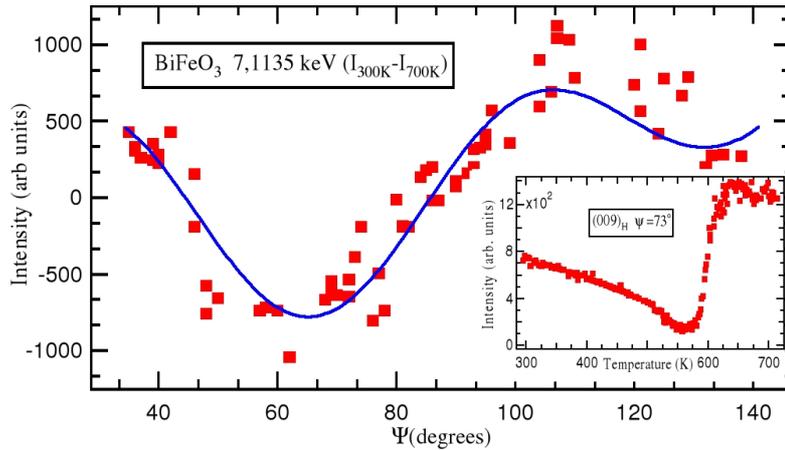

Fig. 5. Difference of two sets of data displayed in figure 4, I(v) – I(v = 0), together with expression (7) derived from our model of electronic structure in bismuth ferrite. Multipoles t, u and v in (7) are set to values derived from fits to data in figure 4, namely, t = + 1.19 ± 0.07, u = − 6.20 ± 0.16 and v = 0.673 ± 0.014. Inset: Temperature dependence of the Bragg spot (0, 0, 9)$_H$ at an azimuthal angle ψ = 73°.

We use the purely real quadrupole (4), with projections Q = ± 1, as a common factor in the final expression for the unit-cell structure factor. The remaining charge-like quadrupole, Q = ± 2, is accounted for in a ratio $\Psi^2_{+2}/\Psi^2_{+1}$ = –it. Calculations using an ideal cycloid show that t is purely real and t = + 1 (Scagnoli and Lovesey.[27]). The contribution from the hexadecapole in (2) is captured by u = 3⟨$T^4_{+3}$⟩'/($\Psi^2_{+1}\sqrt{2}$). In the absorption profile we invoke overlap of the two events, E1-E1 and E2-E2, which occur at different energies with different widths. Lastly, the magnetic contribution to the structure factor is v = $3\langle T^1_{-1} - T^1_{+1}\rangle / (2\Psi^2_{+1})$. Note that t, u and v are all treated as purely real quantities to be inferred from our data. Since t and v both relate to an E1-E1 event they are nothing more than ratios of the appropriate multipoles that we have shown. On the other hand, u has to include a ratio of radial integrals for E2 and E1 events, namely, ($f$ [q $\{R^2\}_{sd}]^2/[\{R\}_{sp}]^2$) where $\{R\}_{sp}$ and $\{R^2\}_{sd}$, respectively, are radial integrals for E1 and E2 events at the K-edge of iron. A multiplicative factor in u, denoted here by $f$, measures the admixture of E1-E1 and E1-E2 events. While $f$ might depend on energy it can be taken purely real without influencing conclusions, because it accompanies the principal harmonic (2).

*E-mail: angelrf86@gmail.com

Incorporating the two types of T & T scattering, expressions (2) and (4), and scattering by magnetic dipoles, we arrive at a unit-cell structure factor that gives an adequate account of all our data,

$F_{\pi'\sigma}$ = t cos$\theta$ sin($\psi$) + u cos$^3\theta$ cos(3$\psi$)–i sin$\theta$ cos(2$\psi$)–v[sin$\theta$–i cos$\theta$sin($\psi$)].   (6)

Writing I(v) = $|F_{\pi'\sigma}(v)|^2$ the difference in intensity at the two temperatures is,
I(v) − I(v = 0) =
  v[v(1 − cos$^2\theta$ cos$^2(\psi)$) − sin2$\theta$ (t sin($\psi$) + u cos$^2\theta$ cos(3$\psi$) + cos(2$\psi$) sin($\psi$))].   (7)

Fits of I(0) to data for T & T scattering displayed in figure 4 (700 K, filled dots) yield values t = + 1.19 ± 0.07, which is close enough to the ideal value to give great confidence, and u = −6.20 ± 0.16. A fit of I(v) to data gathered at 300 K, figure 4 (empty dots), yields v = 0.673 ± 0.014 for the magnetic dipole, with t and u set to aforementioned values. For completeness, we show in figure 5 difference data, taken from figure 4, together with the appropriate expression for intensity (7) evaluated with our estimates of the three multipoles.

We bring our Letter to a close with a survey of our observations and the interpretation we construct. Above the Néel temperature, $T_N$ ≈ 640 K, our azimuthal-angle data are adequately explained by a model with minimal complexity. It includes Templeton and Templeton (T & T) scattering from charge-like quadrupoles and hexadecapoles.[25] A contribution by quadrupoles heralds a new chiral phase, in which quadrupoles participate in a circular cycloid. A test of
chirality in electronic structure is to see whether or not it couples to circular polarization
(helicity) in the x-ray beam
An expression for intensity associated with circular polarization (helicity) in the primary beam, $I_c$, is derived from our unit-cell structure factors [22] and we arrive at,

---
*E-mail: angelrf86@gmail.com

$$I_c = \text{Im.} [(F_{\sigma'\pi})^*(F_{\sigma'\sigma}) + (F_{\pi'\pi})^*(F_{\pi'\sigma})] = 2\cos\theta \cos(\psi) [v \sin\theta - \cos\theta \sin(\psi)]$$
$$\times (t \cos\theta \sin(\psi) + u \cos^3\theta \cos(3\psi) - v \sin\theta). \qquad (8)$$

In expression (8), t and u are charge-like multipoles, which generate T & T scattering, and v is a magnetic dipole absent above $T_N$. Values of the three multipoles, t, u and v, are inferred from data displayed in figure 4, collected above and below the Néel temperature, that are adequately described by $|F_{\pi'\sigma}|^2$ derived from (6). Note that expression (8) does not vanish for v = 0, meaning resonant reflections are affected by circular polarization above $T_N$ with a hitherto unknown phase of the material.

Existence of T & T scattering by quadrupoles in a cycloid implies that the actual chemical structure belongs to an enantiomorphic crystal class lacking a centre of symmetry. Space-group R3 (#146), one of 65 members of the Sohncke sub-group of crystal structures, is a maximal non-isomorphic subgroup of the nominal R3c-group, and thus a likely candidate for a commensurate chiral motif in bismuth ferrite. In which case, a chiral motif and a single domain are implied for the magnetically-ordered state, and this does appear to be the case.[33] A high-quality crystal, from which satellite peaks can be resolved, should show satellite intensity above $T_N$. The domain pattern of propagation vectors should be reproduced on temperature cycling above and below $T_N$ since they are driven by the pre-aligned quadrupoles. This issue could be checked using circular polarized x-rays, due to the helicity properties of this kind of x-rays.

Parallel scenarios merit a mention, e.g., the weak itinerant ferromagnet MnSi, and related materials.[34-36] The compounds use a cubic group $P2_13$ (#198), and exist in both right-handed and left-handed enantiomorphs. A single-valued handedness persists in the ferromagnetic and paramagnetic phases,[37] with chiral fluctuations in MnSi above the Curie temperature observed by inelastic neutron scattering.[38] Notably, a standard example for spontaneous homochirality, sodium chlorate ($NaClO_3$), forms in the chemical structure described by $P2_13$.[39]

MnSi has a Curie temperature $T_c \approx 29.5$ K, and deep in the paramagnetic phase spin fluctuations are isotropic. Perhaps more relevant to the present discussion of bismuth ferrite is another iron-based chiral magnet. FeGe, iso-structural with MnSi, has a high Curie temperature, $T_c \approx 278.2$ K, and precursor activity is well-established.[36,40] Ferromagnetic spirals have a period $\approx$ 180 Å (MnSi) and $\approx$ 700 Å (FeGe), to be compared with a period $\approx$ 620 Å in bismuth ferrite. On its own, an antisymmetric exchange-interaction (Dzialoshinskii-Moriya) will promote an orthogonal arrangement of spins that can disturb an arrangement of parallel spins, supported by an isotropic Heisenberg exchange plus relatively weak magnetic

---

*E-mail: angelrf86@gmail.com

anisotropy.

Quadrupoles (also higher-order multipoles) as a primary order-parameter is not unusual. However, again, order is achieved at low temperatures, because the underlying mechanism is weak. [21,41-43]

The origin of the charge-like quadrupoles that contribute T & T scattering could be related to bismuth 6s-6p lone pairs, known to drive certain structural distortions. Apart from expected direct hybridization of lone pairs, there is scope for admixture through the agency of oxygen 2p states that contribute to angular anisotropy in valence states observed at iron sites.

Lastly, we examine the possibility that our azimuthal-angle scan at 700 K can be explained by the parity-odd event E1-E2 using the R3c-group, in addition to E2-E2.[12,27] We find polar multipoles, $\langle U^K_Q \rangle$, do not contribute intensity to the $(0, 0, 9)_H$ Bragg spot in channels with unrotated polarization, σ'σ and π'π. The contribution from E1-E2 in the π'σ channel of immediate interest comes from a purely real polar quadrupole, namely, $i(3/\sqrt{5}) \cos^2\theta \langle U^2_0 \rangle$ that is added to the hexadecapole contribution (2). The two contributions to the unit-cell structure factor, E1-E2 plus E2-E2, are in phase quadrature, so there can be no interference between them to lift the pure six-fold periodicity in the E2-E2 contribution to intensity that is lacking in figure 4 (filled dots).

In summary, Bragg diffraction intensities at the nominally forbidden reflection $(0, 0, 9)_H$ of bismuth ferrite, observed below and above the Néel temperature, are consistent with a chiral structure formed by a circular cycloid propagating along $[1, 1, 0]_H$ not previously detected in the paramagnetic phase. The new chiral phase is responsible for some Templeton and Templeton (T & T) scattering at 700 K due to charge fluctuations not contained in the plane of the cycloid. Our extensive sets of azimuthal-angle diffraction data have been used to infer good values of three atomic multipoles involved in the scattering process. A satisfactory minimal model of Fe electronic structure includes a quadrupole (E1-E1 event) and a hexadecapole (E2-E2 event) contributing T & T scattering, plus a magnetic dipole (E1-E1).

**Acknowledgments**

We acknowledge the Diamond Light Source for the beam-time allocation on I16. We have benefited from discussions and correspondence on the question of normalization of our data with Dr K Finkelstein. One of us (SWL) is grateful to Dr D D Khalyavin and Dr K S Knight for valuable discussion about the explanation of results offered in the communication. We are also grateful with G Catalan, who has provided the single crystal for performing the experiment. Financial support has been received from Spanish FEDER-MiCiNN Grant No. Mat2011-27573-C04-02. One of us (ARF) is grateful to Gobierno del Principado de Asturias for the financial support from Plan de Ciencia, Tecnología e innovación (PTCI) de Asturias. We thank Diamond Light Source for access to beamline I16 (MT7720) that contributed to the results presented here.

---

*E-mail: angelrf86@gmail.com

**References**

1) A. M. Kadomtseva, A. K. Zvezdin, Y. F. Popov, A. P. Pyatakov, and G. P. Vorob'ev: J. Exp. Theor. Phys. Lett. **79** (2004) 571.
2) G. Catalan and S. F. Scott: Adv. Mater. **21** (2009) 2463.
3) J. F. Scott: Adv. Mater. **22** (2010) 2106.
4) D. C. Arnold, K. S. Knight, G. Catalan, S. A. T. Redfern, J. F. Scott, P. Lightfoot, and F. D. Morrison: Adv. Funct. Mater. **20** (2010) 2116.
5) D. Lebeugle, D. Colson, A. Forget, M. Viret, A. M. Bataille, and A. Gukasov: Phys. Rev. Lett. **100** (2008) 227602.
6) S. Lee, T. Choi, W. Ratcliff, R. Erwin, S. W. Cheong, and V. Kiryukhin: Phys. Rev. B **78** (2008) 100101.
7) F. de Bergevin and M. Brunel: Acta Crystallogr. Sect. **37** (1981) 314.
8) D. Gibbs, G. Grubel, D. R. Harshman, E. D. Isaacs, D. B. McWhan, D. Mills, and C. Vettier: Phys. Rev. B **43** (1991) 5663.
9) S. W. Lovesey: Rep. Prog. Phys. **56** (1993) 25.
10) H. C. Walker, F. Fabrizi, L. Paolasini, F. de Bergevin, J. Herrero-Martin, A. T. Boothroyd, D. Prabhakaran, and D. F. McMorrow: Science **333** (2011) 1273.
11) V. E. Dmitrienko, K. Ishida, A. Kirfel, and E. N. Ovchinnikova: Acta Crystallogr. **A61** (2005) 481.
12) S. W. Lovesey, E. Balcar, K. S. Knight, and J. Fernández-Rodriguez: Phys. Rep. **411** (2005) 233.
13) S. W. Lovesey and E. Balcar: J. Phys. Soc. Japan **82** (2013) 021008.
14) S. W. Lovesey and V. Scagnoli: J. Phys.: Condens. Matter **21** (2009) 474214.
15) J. Fernandez-Rodriguez, J. A. Blanco, P. J. Brown, K. Katsumata, A. Kikkawa, F. Iga, and S. Michimura: Phys. Rev. B **72** (2005) 052407.
16) S. W. Lovesey, J. Fernandez-Rodriguez, J. A. Blanco, D. S. Sivia, K. S. Knight, and L. Paolasini: Phys. Rev. B **75** (2007) 014409.
17) J. Fernandez-Rodriguez, V. Scagnoli, C. Mazzoli, F. Fabrizi, S. W. Lovesey, J. A. Blanco, D. S. Sivia, K. S. Knight, F. de Bergevin, and L. Paolasini: Phys. Rev. B **81** (2010) 085107.
18) S.B Wilkins, R. Caciuffo, C. Detlefs, J. Rebizant, E. Colineau, F. Wastin, and G. H. Lander: Phys. Rev. B **73** (2006) 060406.
19) K. Finkelstein, Q. Shen, and S. Shastri: Phys. Rev. Lett. **69** (1992) 1612.
20) S. W. Lovesey and K. S. Knight: J. Phys.: Condens. Matter **12** (2000) 2367.

*E-mail: angelrf86@gmail.com

21) Y. Tanaka, T. Inami, S. W. Lovesey, K. S. Knight, F. Yakhou, D. Mannix, J. Kokubun, M. Kanazawa, K. Ishida, S. Nanao, T. Nakamura, H. Yamauchi, H. Onodera, K. Ohoyama, and Y. Yamaguchi: Phys. Rev. B **69** (2004) 024417.
22) J. Fernandez-Rodriguez, S. W. Lovesey, and J. A. Blanco: Phys. Rev. B **77** (2008) 094441.
23) S. W. Lovesey, A. Rodriguez-Fernandez and J. A. Blanco: Phys. Rev. B **83** (2011) 054427.
24) R. Palai, R. S. Katiyar, H. Schmid, P. Tissot, S. J. Clark, J. Robertson, S. A. T. Redfern, G. Catalan, and J. F. Scott: Phys. Rev. B **77** (2008) 014110.
25) A. Palewicz, I. Sosnowska, R. Przeniosło and A. Hewat: Acta Physica Polonica A **117** (2010) 296.
26) D. H. Templeton and L. K. Templeton: Acta Crystallogr. **41** (1985) 365; ibid **42** (1986) 478.
27) V. Scagnoli and S. W. Lovesey: Phys. Rev. B **79** (2009) 035111.
28) J. P. Hannon, G. T. Trammell, M. Blume, and D. Gibbs: Phys. Rev. Lett. **61** (1988) 1245; ibid **62** (1989) 2644 (E).
29) R. Przeniosło, M. Regulski, and I. Sosnowska: J. Phys. Soc. Japan **75** (2006) 084718.
30) S. W. Lovesey and E. Balcar: J. Phys.: Condens. Matter **9** (1997) 8679.
31) P. Carra, B. T. Thole, M. Altarelli, and X. Wang: Phys. Rev. Lett. **70** (1993) 694.
32) S. W. Lovesey: J. Phys.: Condens. Matter **10** (1998) 2505.
33) R. D. Johnson, P. Barone, A. Bombardi, R. J. Bean, S. Picozzi, P. G. Radaelli, Y. S. Oh, S. W. Cheong, and L. C. Chapon: Phys. Rev. Lett. **110** (2013) 217206.
34) M. Ishida, Y. Endoh, S. Mitsuda, Y. Ishikawa, and M. Tanaka: J. Phys. Soc. Japan **54** (1985) 2975.
35) V. A. Dyadkin, S. V. Grigoriev, D. Menzel, D. Chernyshov, V. Dmitriev, J. Schoenes, S. V. Maleyev, E. V. Moskvin, and H. Eckerlebe: Phys. Rev. B **84** (2011) 014435.
36) H. Wilhelm, M. Baenitz, M. Schmidt, C. Naylor, R. Lortz, U. K. Rössler, A. A. Leonov, and A. N. Bogdanov: J. Phys.: Condens. Matter **24** (2012) 294204.
37) V. Dmitriev, D. Chernyshov, S. Grigoriev, and V. Dyadkin: J. Phys.: Condens. Matter **24** (2012) 366005.
38) B. Roessli, P. Böni, W. E. Fischer, and Y. Endoh: Phys. Rev. Lett. **88** (2002) 237204.
39) C. Viedma and P. Cintas: Chem. Commun. **47** (2011) 12786.
40) E. Moskvin, S. Grigoriev, V. Dyadkin, H. Eckerlebe, M. Baenitz, M. Schmidt, and H. Wilhelm: Phys. Rev. Lett. **110** (2013) 077207.
41) P. Morin, D. Schmitt, and E. de Lacheisserie: J. Magn. Magn. Mater. **30** (1982) 257.
42) T. Sakakibara, T. Tayama, T. Onimaru, D. Aoki, Y. Onuki, H. Sugawara, Y. Aoki, and H. Sato: J. Phys.: Condens. Matter **15** (2003) S2055.
43) Y. Kuramoto, H. Kusunose, and A. Kiss: J. Phys. Soc. Japan **78** (2009) 072001.

*E-mail: angelrf86@gmail.com

44) J. Kokubun, A. Watanabe, M. Uehara, Y. Ninomiya, H. Sawai, N. Momozawa, K. Ishida, and V. E. Dmitrienko: Phys. Rev. B **78** (2008) 115112.
45) S. W. Lovesey, V. Scagnoli, M. Garganourakis, S. M. Koohpayeh, C. Detlefs, and U. Staub: J. Phys.: Condens. Matter **25** (2013) 362202.

*E-mail: angelrf86@gmail.com